\documentclass[11pt,a4paper]{article}
\usepackage{a4wide}
\usepackage{amssymb}
\usepackage{amsmath}
\usepackage{graphicx}
\usepackage{wrapfig}
\usepackage{caption}
\usepackage{subcaption}
\usepackage{bbm}
\usepackage{cite}
\usepackage{hyperref}

\newcommand{\be}{\begin{equation}}
\newcommand{\ee}{\end{equation}}
\newcommand{\bea}{\begin{eqnarray}}
\newcommand{\eea}{\end{eqnarray}}

\newcommand{\bdm}{\begin{displaymath}}
\newcommand{\edm}{\end{displaymath}}

\begin{document}
{\renewcommand{\thefootnote}{\fnsymbol{footnote}}
\begin{titlepage}

\noindent
\begin{center}
\vspace*{1cm}

{\Large \bf  Constraints on the duration of inflation \\from entanglement entropy bounds}

\vskip 2cm

{\bf Suddhasattwa Brahma}\footnote{\tt suddhasattwa.brahma@gmail.com} 
\vskip 1cm
Higgs Centre for Theoretical Physics, School of Physics \& Astronomy,
\\ University of Edinburgh, Edinburgh EH9 3FD, UK.\\[3mm]
Department of Physics, McGill University, Montr\'eal, QC H3A 2T8, Canada. 
\vspace{2cm}

\begin{abstract}
Using the fact that we only observe those modes which exit the Hubble horizon during inflation, one can calculate the entanglement entropy of such long-wavelength perturbations by tracing out unobservable sub-Hubble fluctuations they are coupled to. On requiring that this perturbative entanglement entropy, which increases with time, obey the covariant entropy bound for an accelerating background, we find an upper bound on the duration of inflation. This presents a new perspective on the (meta-)stability of de Sitter spacetime and an associated lifetime for it. 
\end{abstract}

\vskip 3cm

\end{center}

\end{titlepage}
\setcounter{footnote}{0}
{\renewcommand{\thefootnote}{\arabic{footnote}}


Although cosmic inflation is widely regarded as the standard paradigm for the early universe, its embedding into a fundamental theory of quantum gravity (QG) remains an open question. Recently, there have been different arguments against long-lived accelerating spacetimes, especially in the context of string theory (ST) \cite{Palti:2019pca, Ooguri:2018wrx}. One such conjecture states that trans-Planckian modes should never cross the Hubble horizon during inflation, leading to an upper bound on the number of $e$-foldings \cite{Bedroya:2019snp}:
\begin{eqnarray}\label{TCC}
	N < \ln\left(\frac{M_{\rm Pl}}{H}\right)\,,
\end{eqnarray}
where $H$ denotes the Hubble parameter during inflation. Although the physical motivation behind this conjecture -- a trans-Planckian mode should never become part of late-time macroscopic inhomogeneities -- has been heavily debated \cite{Dvali:2020cgt, Burgess:2020nec}, it does find some connections to other aspects of the ST `swampland' \cite{Bedroya:2020rmd, Aalsma:2020aib, Andriot:2020lea, Berera:2020dvn, Brahma:2019vpl, Rudelius:2021oaz}. In particular, a corollary of this is that only extremely short-lived de Sitter (dS) spaces can arise in a UV-consistent theory \cite{Bedroya:2019snp, Ooguri:2018wrx}.

Indeed, it has been long argued that dS space is meta-stable from different points of view\footnote{These arguments, to name a few, are based on the finiteness of dS entropy \cite{Goheer:2002vf, ArkaniHamed:2007ky}, the typical lifetime of $4$-dimensional dS vacua in ST \cite{Kachru:2003aw}, non-perturbative effects in the context of eternal inflation \cite{Dubovsky:2008rf} or treating dS as a coherent state \cite{Dvali:2017eba, Brahma:2020tak, Brahma:2020htg} and so on.}. There are three time-scales which are often associated with the lifetime of dS -- the scrambling time $\sim H^{-1}\, \ln(S_{\rm dS})$ \cite{Susskind:2011ap} corresponding to \eqref{TCC}, the quantum breaking time $\sim H^{-1}\,S_{\rm dS}$ and the Poincar\'e recurrence time $\propto e^{S_{\rm dS}}$, where the Gibbons-Hawking entropy for dS is given by $S_{\rm dS} \sim \left(M_{\rm Pl}/H\right)^2$ \cite{Gibbons:1977mu}. Clearly, \eqref{TCC} puts an upper bound on the number of $e$-foldings that is much smaller than the other two time-scales, with drastic implications for inflation \cite{Bedroya:2019tba}.

In this essay, we present a different argument for finding the maximum amount of $e$-foldings allowed for inflation, and therefore, set an upper bound on the lifetime of dS. Instead of invoking any QG reasoning, we employ a bottom-up argument by requiring that the entanglement entropy (EE) of scalar perturbations during inflation be bounded by the Gibbons-Hawking entropy. We note that the first arguments in favor of the so-called dS conjecture also followed from an application of the covariant entropy bound (CEB) and the distance conjecture \cite{Ooguri:2018wrx}. However, that derivation was \textit{i)} intimately tied to details of ST and \textit{ii)} valid only in asymptotic regions of moduli space. Here, we circumvent both these obstructions.

Discussions of EE have become ubiquitous in the context of gravity. However, in most cases, one considers the EE  between different geometric regions of space -- in the context of black holes \cite{Solodukhin:2011gn}, Minkowski \cite{Casini:2009sr} or dS space \cite{Maldacena:2012xp}. Nevertheless, it is not necessary to define a subsystem which is separated out in the position-space domain, \textit{e.g.} demarcation by a black hole horizon. In cosmology, it is more instructive to consider EE between different bands in momentum-space since it is the correlation functions of the momentum modes of cosmological perturbations which are generally probed. For momentum-space, vacuum of the free field theory is factorized, and therefore, any EE come from interactions which lead to mode-coupling. 

One can calculate the perturbative EE in momentum-space for a scalar in flat spacetime as outlined, say, in \cite{Balasubramanian:2011wt}. The full Hilbert space can be partitioned into two parts separated by some fiducial momentum scale $\mu$ such that $\mathcal{H} = \mathcal{H}_{\mathcal{S}} \otimes \mathcal{H}_{\mathcal{E}}$. The Hamiltonian of the system is decomposed as
\begin{eqnarray}
	H = H_{\mathcal{S}}\otimes \mathbbm{1} + \mathbbm{1}\otimes H_{\mathcal{E}} + \lambda H_{\rm int}\,,
\end{eqnarray}
where $H_{\mathcal{E, S}}$ are the free Hamiltonians of the respective subsystems and the interacting Hamiltonian $H_{\rm int}$ has a coupling parameter $\lambda$. The ground state is the product of the individual harmonic vacua of $H_A$ and $H_B$, \textit{i.e.} $|0,0\rangle = |0\rangle_{\mathcal{S}} \otimes |0\rangle_{\mathcal{E}}$. The energy eigenbasis of $\mathcal{S}$ and $\mathcal{E}$ are denoted by $|n\rangle$ and $|N\rangle$, respectively, while the corresponding energy eigenvalues by $E_n$ and $\bar{E}_N$. The (perturbative) interacting vacuum can be written as (up to normalization)
\begin{eqnarray}
	|\Omega\rangle = |0,0\rangle + \sum\limits_{n\neq 0} A_n |n,0\rangle + \sum\limits_{n\neq 0} B_N |0,N\rangle +\sum\limits_{n,N\neq 0} C_{n,N} |n,N\rangle\,,
\end{eqnarray}
where the matrix elements $A_n, B_N, C_{n,N}$ are calculated using standard perturbation theory. The reduced density matrix corresponding to subsystem $\mathcal{S}$ is obtained by tracing out the $\mathcal{E}$ modes. From that, one can extract the leading order contribution to the (von Neumann) EE:
\begin{eqnarray}\label{Flat_EE}
S_{\rm ent} = - \lambda^2 \log\left(\lambda^2\right) \sum\limits_{n,N\neq 0}\; \dfrac{|\langle n, N| H_{\rm int} |0,0\rangle|^2}{(E_0 + \tilde{E}_0 - E_n - \tilde{E}_N)^2} + \mathcal{O}(\lambda^2)\,,
\end{eqnarray} 
where it is understood that at least one momentum in the matrix element is below $\mu$ and at least one momentum is above.

The main calculation of \cite{Brahma:2020zpk} was to extend this result to an inflating background, which we now present. Considering density perturbations in the comoving gauge 
\begin{eqnarray}
 {\rm d} s^2= -a^2(\tau) \left[{\rm d}\tau^2 - \left(1+2\zeta\right){\rm d} x^2\right]\,,
\end{eqnarray}
where $a$ and $\epsilon$ are the usual symbols for the scale factor and the (first) slow-roll parameter for a quasi-dS expansion written as functions of the conformal time $\tau$. The quadratic action for $\zeta$, in terms of momentum modes, describes a collection of harmonic oscillators with a time-dependent mass term. This implies a difference between quasi-dS geometry and Minkowski background \cite{Brahma:2020zpk} -- sub-Hubble modes ($k \gg aH$) are in their quantum (Bunch-Davies) vacuum  while the super-Hubble ones ($k\ll aH$) are in a (two-mode) squeezed state $\left|SQ\left(k,\tau\right) \right\rangle$ \cite{Albrecht:1992kf}. Thus, the ground state factorizes as
\begin{eqnarray}\label{Ground_state}
	|0,0\rangle = |0\rangle_{\mathcal{E}: k>aH} \otimes |SQ\rangle_{\mathcal{S}: k<aH}\,.
\end{eqnarray}
Since the modes which exit the horizon during inflation can only be later observed, we consider the super-Hubble modes as our system ($\mathcal{S}$) while the sub-Hubble ones as environment ($\mathcal{E}$) (Fig.~\ref{Fig:1}). (The dynamics of this system has been studied in \cite{Brahma:2021mng}.) Finally, the non-linearity of GR also provides us with an interaction term which couples the sub- and super-Hubble modes, and thus, this interaction is universal and can never be \textit{turned off}. 

\begin{figure}[h]
	\centering
	\includegraphics[width=0.3\textwidth]{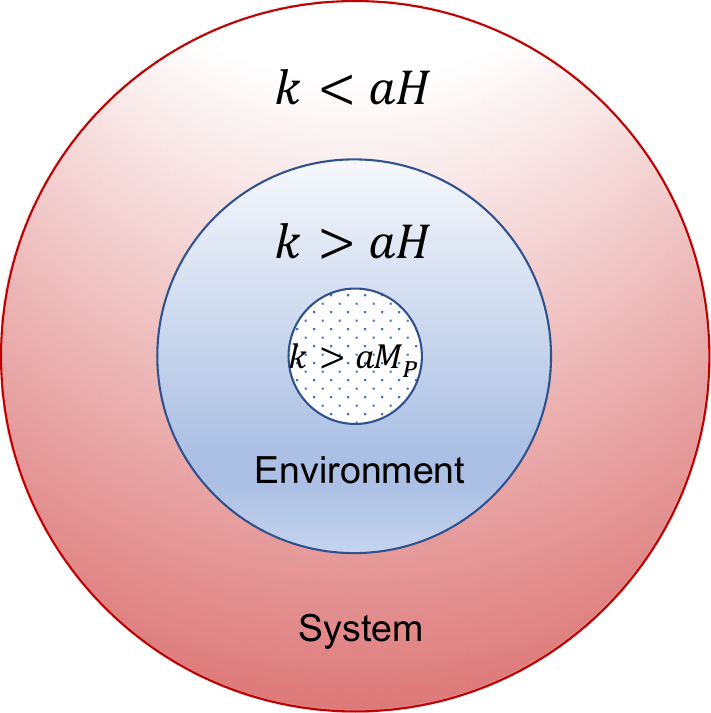}
	\caption{Schematic illustration of the system and environment modes for this setup. Gravity plays the role of providing a natural scale -- the comoving Hubble horizon -- which demarcates ``long'' and ``short'' degrees of freedom (dofs). We impose the Planck mass as the natural cutoff for the UV-modes and assume that these can be properly accounted for within some QG theory.}\label{Fig:1}
\end{figure}

Mathematically, this translate into having a Hilbert space: $\mathcal{H} = \mathcal{H}_{\mathcal{S}} \otimes \mathcal{H}_{\mathcal{E}}$ where $\mathcal{H}_{\mathcal{S}} = \prod_k \mathcal{H}_{k}$, \;\;$|k|< aH$ and similarly for the sub-Hubble modes. The full Hamiltonian is given by $H = H^{(2)}_{\mathcal{S}} + H^{(2)}_{\mathcal{E}} + H_{\rm int}$, where
\begin{eqnarray}
 H_{\rm int} \, = \, \frac{M_{\rm Pl}^2}{2}\int d^3x \; \epsilon^2\, a \,\zeta (\partial \zeta)^2\,,
\end{eqnarray} 
is the leading order cubic non-Gaussian term (since $\zeta$ ``freezes'' outside the horizon) out of all the available interactions \cite{Maldacena:2002vr}. We now need to apply time-dependent perturbation theory to calculate the matrix elements since $\lambda= \sqrt{\epsilon}/(2\sqrt{2} a M_{\rm Pl})$ is time-dependent. Also, there is no well-defined notion of energy for the squeezed state but luckily we only need energy differences in \eqref{Flat_EE}, which can be expressed in terms of the physical momenta. Using these inputs, one can carefully evaluate \eqref{Flat_EE} for inflationary perturbations \cite{Brahma:2020zpk}, resulting in the EE (per unit physical volume):
\begin{eqnarray}\label{EE}
s_{\rm ent} \sim \epsilon\; H^2 \;M_{\rm Pl}\; (a_f/a_i)^2\,,
\end{eqnarray}
where $(a_f)$ $a_i$ is the scale factor at the (end) beginning of inflation. In this calculation, it was assumed that there were no super-Hubble modes at the beginning of inflation and both $(H, \epsilon)$ remain constant. It was shown that the dominant contribution comes from the squeezed configuration, and in the large squeezing limit, we present the leading order estimate omitting some $\mathcal{O}(1)$ factors as well as small logarithmic corrections \cite{Brahma:2020zpk}. 

Firstly, we note that the resulting EE between the sub- and super-Hubble modes is sensitive to the UV-cutoff $M_{\rm Pl}$, as expected. However, what is remarkable about this result is that the EE increases secularly with time, as signified by the $(a_f/a_i)^2$ term. The intuitive reason for this is that the dimension of $\mathcal{H}_\mathcal{S}$ increases with time as modes get stretched outside the horizon. Although the EE here is a perturbative result for density fluctuations, given enough time, it can overcome the background Gibbons-Hawking entropy. Growth of entropy often leads to deep puzzles in theoretical physics and we make our most crucial observation in this context. We require that the \textit{total EE obeys the CEB} \cite{Bousso:1999xy}, \textit{i.e.} the EE in a Hubble patch can, at most, saturate the entropy corresponding to the apparent horizon ($S_{\rm dS}$). This implies
\begin{eqnarray}
	\epsilon\;  e^{2N} < \left(\frac{M_{\rm Pl}}{H}\right) \;\;\;\; \Rightarrow \;\; N < \frac{1}{2} \ln \left(\frac{M_{\rm Pl}}{H}\right) - \frac{1}{2} \ln \epsilon\,.
\end{eqnarray}
Using that the observed power spectrum $P_{\zeta} \sim 10^{-9}$, one finds $\epsilon \sim 10^{9} \left(H/M_{\rm Pl}\right)^2$ and therefore,
\begin{eqnarray}\label{Final}
	N < \frac{3}{2} \ln\left(\frac{M_{\rm Pl}}{H}\right) - \frac{9}{2}\ln 10\,.
\end{eqnarray}
This bound on the number of $e$-foldings is very similar to the one in \eqref{TCC}, without requiring \textit{any} QG input, and predicts an upper limit on the lifetime of dS given by
\begin{eqnarray}\label{dS_lifetime}
	T< \frac{1}{H}\left[\frac{3}{2} \ln\left(\frac{M_{\rm Pl}}{H}\right) - \frac{9}{2}\ln 10\right]\,,
\end{eqnarray}
closely related to the scrambling time up to small factors. Our result has far-reaching implications both for UV-completions of dS space as well as phenomenological predictions of inflation\footnote{
A direct evaluation of EE in momentum space for a scalar field on pure dS, as well as a more sophisticated calculation of the inflationary system allowing for a slowly varying $H$ and $\epsilon$, shall be carried out in the future \cite{us}.}.

To summarize, for a dS geometry, an observer has access to only part of the entire spacetime. In particular for inflation, tracing out the unobservable sub-Hubble modes leads to a non-zero EE for the curvature perturbations that increases with time. However, since the EE can, at best, saturate the CEB, this puts on an upper limit on the duration of inflation. Our calculation provides an universal limit since we take the simplest case of a minimally-coupled scalar field -- any additional fields or extra-couplings (which give rise to stronger non-Gaussianity) would only enhance the EE and strengthen our result. We emphasize that our bound does not arise from demanding a finite-dimensional Hilbert space for dS \cite{Banks:2000fe} or that we live in an asymptotically dS universe \cite{Banks:2003pt}. Finally, note that EE for other early-universe scenarios do not produce such bounds on the lifetime \cite{Brahma:2020rtx}.

\vspace{3mm}
\noindent {\bf Acknowledgements:} I am grateful to Robert Brandenberger for multiple discussions. \\ 
SB is supported in part by the Higgs Fellowship.




\begin{thebibliography}{99}


\bibitem{Palti:2019pca}
E.~Palti, 
Fortsch. Phys. \textbf{67}, no.6, 1900037 (2019)
[arXiv:1903.06239 [hep-th]].

\bibitem{Ooguri:2018wrx}
H.~Ooguri, E.~Palti, G.~Shiu and C.~Vafa,
Phys. Lett. B \textbf{788}, 180-184 (2019)
[arXiv:1810.05506 [hep-th]].

\bibitem{Bedroya:2019snp}
A.~Bedroya and C.~Vafa,
JHEP \textbf{09}, 123 (2020)
[arXiv:1909.11063 [hep-th]].

\bibitem{Burgess:2020nec}
C.~P.~Burgess, S.~P.~de Alwis and F.~Quevedo,
[arXiv:2011.03069 [hep-th]].

\bibitem{Dvali:2020cgt}
G.~Dvali, A.~Kehagias and A.~Riotto,
[arXiv:2005.05146 [hep-th]].

\bibitem{Bedroya:2020rmd}
A.~Bedroya,
[arXiv:2010.09760 [hep-th]].

\bibitem{Rudelius:2021oaz}
T.~Rudelius,
[arXiv:2101.11617 [hep-th]].

\bibitem{Andriot:2020lea}
D.~Andriot, N.~Cribiori and D.~Erkinger,
JHEP \textbf{07}, 162 (2020)
[arXiv:2004.00030 [hep-th]].

\bibitem{Berera:2020dvn}
A.~Berera, S.~Brahma and J.~R.~Calder\'on,
JHEP \textbf{08}, 071 (2020)
[arXiv:2003.07184 [hep-th]].

\bibitem{Brahma:2019vpl}
S.~Brahma,
Phys. Rev. D \textbf{101}, no.4, 046013 (2020)
[arXiv:1910.12352 [hep-th]].

\bibitem{Aalsma:2020aib}
L.~Aalsma and G.~Shiu,
JHEP \textbf{05}, 152 (2020)
[arXiv:2002.01326 [hep-th]].

\bibitem{Goheer:2002vf}
N.~Goheer, M.~Kleban and L.~Susskind,
JHEP \textbf{07}, 056 (2003)
[arXiv:hep-th/0212209 [hep-th]].

\bibitem{ArkaniHamed:2007ky}
N.~Arkani-Hamed, S.~Dubovsky, A.~Nicolis, E.~Trincherini and G.~Villadoro,
JHEP \textbf{05}, 055 (2007)
[arXiv:0704.1814 [hep-th]].

\bibitem{Kachru:2003aw}
S.~Kachru, R.~Kallosh, A.~D.~Linde and S.~P.~Trivedi,
Phys. Rev. D \textbf{68}, 046005 (2003)
doi:10.1103/PhysRevD.68.046005
[arXiv:hep-th/0301240 [hep-th]].

\bibitem{Dubovsky:2008rf}
S.~Dubovsky, L.~Senatore and G.~Villadoro,
JHEP \textbf{04}, 118 (2009)
[arXiv:0812.2246 [hep-th]].

\bibitem{Dvali:2017eba}
G.~Dvali, C.~Gomez and S.~Zell,
JCAP \textbf{06}, 028 (2017)
[arXiv:1701.08776 [hep-th]].

\bibitem{Brahma:2020tak}
S.~Brahma, K.~Dasgupta and R.~Tatar,
JHEP \textbf{02}, 104 (2021)
[arXiv:2007.11611 [hep-th]].

\bibitem{Brahma:2020htg}
S.~Brahma, K.~Dasgupta and R.~Tatar,
JHEP \textbf{07}, 114 (2021)
[arXiv:2007.00786 [hep-th]].

\bibitem{Susskind:2011ap}
L.~Susskind,
[arXiv:1101.6048 [hep-th]].

\bibitem{Gibbons:1977mu}
G.~W.~Gibbons and S.~W.~Hawking,
Phys. Rev. D \textbf{15}, 2738-2751 (1977)

\bibitem{Bedroya:2019tba}
A.~Bedroya, R.~Brandenberger, M.~Loverde and C.~Vafa,
Phys. Rev. D \textbf{101}, no.10, 103502 (2020)
[arXiv:1909.11106 [hep-th]].


\bibitem{Solodukhin:2011gn}
S.~N.~Solodukhin,
Living Rev. Rel. \textbf{14} (2011), 8
[arXiv:1104.3712 [hep-th]].

\bibitem{Casini:2009sr}
H.~Casini and M.~Huerta,
J. Phys. A \textbf{42} (2009), 504007
[arXiv:0905.2562 [hep-th]].

\bibitem{Maldacena:2012xp}
J.~Maldacena and G.~L.~Pimentel,
JHEP \textbf{02} (2013), 038
[arXiv:1210.7244 [hep-th]].

\bibitem{Balasubramanian:2011wt}
V.~Balasubramanian, M.~B.~McDermott and M.~Van Raamsdonk,
Phys. Rev. D \textbf{86} (2012), 045014
[arXiv:1108.3568 [hep-th]].

\bibitem{Brahma:2020zpk}
S.~Brahma, O.~Alaryani and R.~Brandenberger,
Phys. Rev. D \textbf{102} (2020) no.4, 043529
[arXiv:2005.09688 [hep-th]].

\bibitem{Albrecht:1992kf}
A.~Albrecht, P.~Ferreira, M.~Joyce and T.~Prokopec,
Phys. Rev. D \textbf{50} (1994), 4807-4820
[arXiv:astro-ph/9303001 [astro-ph]].

\bibitem{Brahma:2021mng}
S.~Brahma, A.~Berera and J.~Calder\'on-Figueroa,
[arXiv:2107.06910 [hep-th]].

\bibitem{Maldacena:2002vr}
J.~M.~Maldacena,
JHEP \textbf{05} (2003), 013
[arXiv:astro-ph/0210603 [astro-ph]].

\bibitem{Bousso:1999xy}
R.~Bousso,
JHEP \textbf{07} (1999), 004
[arXiv:hep-th/9905177 [hep-th]].

\bibitem{us}
S. Brahma,
\textit{In preparation}.

\bibitem{Banks:2000fe}
T.~Banks,
Int. J. Mod. Phys. A \textbf{16}, 910-921 (2001)
[arXiv:hep-th/0007146 [hep-th]].

\bibitem{Banks:2003pt}
T.~Banks and W.~Fischler,
[arXiv:astro-ph/0307459 [astro-ph]].


\bibitem{Brahma:2020rtx}
S.~Brahma, R.~Brandenberger and Z.~Wang,
JCAP \textbf{03}, 094 (2021)
[arXiv:2009.12653 [hep-th]].

\end{thebibliography}
\end{document}